\documentclass[twocolumn,aps,prc,superscriptaddress,showpacs]{revtex4}

\usepackage{amssymb}
\usepackage{amsmath}
\usepackage{graphicx}
\usepackage[normalem]{ulem}
\usepackage{multirow}
\usepackage{CJK}
\usepackage[usenames]{color}

\begin{document}


\title{Effect of in-medium nucleon-nucleon cross section on proton-proton momentum correlation in intermediate energy heavy-ion collision}

\author{Ting-Ting Wang}
\affiliation{Shanghai Institute of Applied Physics, Chinese Academy of Sciences, Shanghai 201800, China}
\affiliation{University of Chinese Academy of Sciences, Beijing 100049, China}

\author{Yu-Gang Ma\footnote{Corresponding author: ygma@sinap.ac.cn}}
\affiliation{Shanghai Institute of Applied Physics, Chinese Academy
of Sciences, Shanghai 201800, China}
\affiliation{University of Chinese Academy of Sciences, Beijing 100049, China}
\affiliation{ShanghaiTech University, Shanghai 200031, China}

\author{Chun-Jian Zhang}
\author{Zheng-Qiao Zhang}
\affiliation{Shanghai Institute of Applied Physics, Chinese Academy of Sciences, Shanghai 201800, China}
\affiliation{University of Chinese Academy of Sciences, Beijing 100049, China}

\date{\today}

\begin{abstract}

The proton-proton momentum correlation function  from different rapidity regions are systematically investigated for the Au + Au collisions at different impact parameters and different energies from 400$A$ MeV to 1500$A$ MeV in the framework of the isospin-dependent quantum molecular dynamics model complemented by the $Lednick\acute{y}$ and $Lyuboshitz$ analytical method. In particular, in-medium nucleon-nucleon cross section dependence of the correlation function is brought into focus, while the impact parameter and energy dependence of the momentum correlation function are also explored. The sizes of the emission source are extracted by fitting the momentum correlation functions using the Gaussian source method. We find that the in-medium nucleon-nucleon cross section  obviously influence the proton-proton momentum correlation function which is from the whole rapidity or projectile/target rapidity region at smaller impact parameters, but there is no effect on the mid-rapidity proton-proton momentum correlation function, which indicates that the emission mechanism differs between projectile/target rapidity and mid-rapidity protons.

\end{abstract}
\pacs{25.70.Mn, 24.10.-i, 25.70.Pq, 27.80.+w}

\maketitle

\section{Introduction}
The Hanbury Brown  and Twiss  (HBT) effect was firstly discussed in radio and stellar astronomy. The method was applied to measure the angular diameter and the size of stars by Hanbury Brown  and Twiss \cite{HBT}. Later on, the technique was introduced to research particle physics in 1960s by Goldhaber $et$ $al.$. They studied the angular distribution of identical pion pairs in proton-antiproton annihilations and observed an enhancement of pairs at small relative momenta \cite{Gold}. In the last decade, there has brought great strides in experiments and a larger number of theoretical researches ranging from low energy to high energy HICs \cite{Boal,Heinz1999}. It is well known that the two-particle correlation is sensitive to characteristics of the particle emission source. Recently, the two particle correlation in subatomic physics has been taken as a probe for the space-time geometry of the particle emission source. Correlation between two protons was measured by several experiments and explored by different models. Not only protons, but also the composite light fragments/particles, which will not be discussed in this paper, are also used to carry information on the emission source \cite{Orr,Cao2012}. Due to the rapid development of radioactive nuclear beams, the HBT method is also used to study the exotic structure of nuclei. For instance, there have been several measurements for revealing exotic structure of the neutron-rich nuclei such as $^{6}\textrm{He}$, $^{11}\textrm{Li}$, $^{14}\textrm{Be}$ \cite{Ieki1993,Yamashita2005,Kohley2013} and of proton-rich nuclei such as $^{23}\textrm{Al}$ \cite{ZhouP} as well as $^{22}\textrm{Mg}$ \cite{Ma2015,Fang2016}. In addition, the dependence of the proton-neutron correlation on the binding energy  was also theoretically explored \cite{Wei2004}. Besides the applications of the HBT method to investigate the exotic structure, it has become an important tool in heavy-ion collisions at wide energy range \cite{Bauer1992,Sullivan1993,Pratt1987,Boal,Heinz1999}. For example, in relativistic  energy region, the Collaborations at the RHIC  and the LHC have carried out a lot of experimental measurements of the correlation function of two-pion as a function of energy and system size
\cite{STAR2001,ALICE2011}. What's more,  the same method has been applied to make the first measurement of two-antiproton interaction by analyzing momentum correlation function between antiprotons, namely the quantitative extraction of the scattering length and the effective range which are two key parameters to characterize the strong interaction  for the antiproton interactions  by the STAR Collaboration \cite{zzqnature,Zhang_PhD,Zhang_NSR,Zhang_NST}. Theoretically, the  correlation functions between two identical pions or kaons were also investigated in some simulation work such as the hydrodynamic model  and the AMPT model etc \cite{Heinz1999,Zhang2014,LIN2002,Lisa,Lisa2,ZhangWN}. In intermediate energy region, the two-proton correlation functions have been mostly applied to extract the space-time properties such as the source size and emission time in the nuclear reaction \cite{Ghetti2003}. In addition, there are many investigations on the dependences of the correlation functions in the experiments and theories, such as on  impact parameter \cite{Gong1991,Ma2006},  total momentum of nucleon pairs \cite{Colonna},  isospin of the emission source \cite{Ghetti2004},  nuclear symmetry energy \cite{Chen2003}, nuclear equation of state (EOS) \cite{Ma2006}, the density distribution of valence neutrons in neutron-rich nuclei \cite{Cao2012} and so on.

The investigation of in-medium nucleon-nucleon scattering is of interest in intermediate energy heavy ion reactions. In this energy domain, nucleus-nucleus collisions provide a unique opportunity to form the compressed nuclear matter with a density up to 2-3 times normal nuclear matter density ($\rho_{0}$). The in-medium nucleon-nucleon cross section (NNCS) has a close relation with the nuclear matter density. Therefore, it is an important component in our model simulations. Recently, the medium effects on nucleon-nucleon cross section have been widely investigated by replacing the NNCS in vacuum with an in-medium one and the various effects have been discussed \cite{Ming2017,Haar1987,Cai1998,Zhang2007}.

The dependence of the two-proton correlation function on the in-medium  NNCS has been briefly studied through CRAB code  in a framework of  IQMD \cite{Ma2006}. Since two-particle correlation function, through final-state interactions and quantum statistical effects, have been shown to be a sensitive probe to the space-time distributions of emitted particle in heavy-ion collisions \cite{Bauer1992}, it is of great interest to investigate the in-medium NN cross section effects on the source evolution. In the present paper, we use another theoretical approach which was proposed by $Lednick\acute{y}$ and $Lyuboshitz$ \cite{lednicky1982} to explore the relationship between the above factors and proton-proton correlation function more details. The two-particle correlation at small relative velocities is sensitive to the space-time characteristics of the production process owning to the effects of quantum statistics and final-state interaction \cite{Lednicky2008,Lednicky2009}. In most proton-proton correlation functions, the HBT strength at 20 MeV/c of the p-p relative momentum is taken as a unique quantity to determine the source size or emission time of two-proton emission \cite{Ma2006}. The proton phase spaces of  Au+Au collisions at the freeze-out time generated by the IQMD model are used as the input for the $Lednick\acute{y}$ and $Lyuboshitz$ code and then the effective source  size
of the source will be  extracted.

The rest of paper is organized as follows. In Sec. II we briefly describe the models and formalism used in the present study, $i.e.$, $Lednick\acute{y}$ and $Lyuboshitz$ analytical formalism and an isospin-dependent quantum molecular dynamics model (IQMD) model.  The detailed analysis and discussion of systematic proton-proton momentum correlation function ($C_{pp}$) and extracted source size results for different rapidity regions are given in  different in-medium nucleon-nucleon cross sections, different impact parameters, beam energies for Au + Au collisions at  in Sec. III. In addition, we fit the proton $p_{T}$ spectra with the distribution function from the Blast-Wave model and discuss the relationship between proton-proton correlation strength and radial flow velocity. Finally, in Section IV we summarize the results.

\section{FORMALISM AND MODELS}

\subsection{LEDNICK$\acute{Y}$ and LYUBOSHITZ ANALYTICAL FORMALISM}

Firstly, we would like to present a brief review of the theoretical approach which was proposed by $Lednick\acute{y}$ and $Lyuboshitz$ \cite{lednicky1982} for the HBT analysis. The method is based on the principle that correlation function of identical particles when they are emitted at small relative momenta are determined by the effects of quantum-statistical symmetry $\left(QS\right)$ of particles and the final-state interaction $\left(FSI\right)$ \cite{Koonin1977}. In this technique, we assume the particles emitted by independent one-particle point source and the spin independent in the production progress as well as the two-particle interaction. Then we can investigate particle pairs $\left(1,2\right)$ emitted at small relative momenta. Due to the conditions following references \cite{Lednicky1996} , we neglect the effect of FSI in all pairs $\left(1,i\right)$ and $\left(2,i\right)$ except $\left(1,2\right)$. We can see the progress in fig 1.

\begin{figure}[h]
\includegraphics[width=\linewidth]{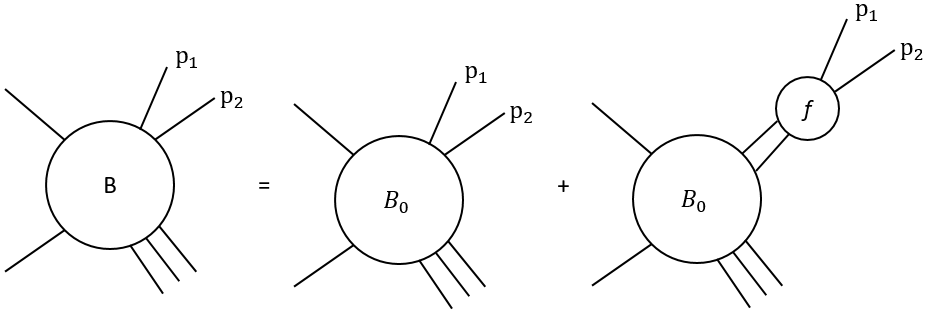}
\caption{The diagram describes the production of interacting particle 1 and 2 \cite{Lednicky1996}.}
\end{figure}

So the correlation function for identical particles takes the expression:

\begin{equation}
B\left(\textbf{p},\textbf{q}\right) = B_{0}\left(\textbf{p},\textbf{q}\right)+B_{1}\left(\textbf{p},\textbf{q}\right),
\end{equation}
where
\begin{equation}
 \textbf{p} = \textbf{p}_{1}+\textbf{p}_{2}, \textbf{q} = \frac{1}{2} | \textbf{p}_{2}-\textbf{p}_{1}|
\end{equation}
are the total momentum and relative momentum of the particle pair, respectively. In Eq. (1),  $B_{0}\left(\textbf{p},\textbf{q}\right)$ is the contribution of quantum statistics effect, described by the formula:
\begin{equation}
 B_{0}\left(\textbf{p},\textbf{q}\right) =  g_{0}\cos\left(\textbf{q}\textbf{x}\right).
\end{equation}
Here $g_{0}$ is the spin factor.

Then, the function $B_{1}\left(\textbf{p},\textbf{q}\right)$ can be expressed through the symmetrizied Bethe-Salpeter amplitude $\psi\left(S\right)$ which can be approximated by the outer region solution of the scattering problem \cite{zzqnature}:
\begin{multline}
B_{1}\left(\textbf{p},\textbf{q}\right)  = \sum_{S}\rho_{S,\textbf{p}}\left(x_{1},x_{2},x_{1}^{,},x_{2}^{,}\right)\times\psi_{\textbf{p},\textbf{q}}^{S}\left(x_{1},x_{2}\right)\\
\times\psi_{\textbf{p},\textbf{q}}^{S}\left(x_{1}^{,},x_{2}^{,}\right)d^{4}x_{1}d^{4}x_{2}d^{4}x_{1}^{,}d^{4}x_{2}^{,},
\end{multline}
with $\rho _{S,\textbf{p}}$ is the two-particle density matrix.

Next, we can introduce the detailed analytical calculation of the proton-proton correlation function \cite{zzqnature}.
The proton-proton correlation function, $C_{pp}\left(\textbf{k}^*,r_{0}\right)$ can be described by the $Lednick\acute{y}$ and $Lyuboshitz$ analytical method \cite{lednicky1982,zzqnature}.
In this model, the space distribution of the Gaussian source was simulated according to the following function:
\begin{equation}
S\left(\textbf{r}^*\right)\approx\exp\left(-\textbf{r}^{*^{2}}/\left(4r_{0}^2\right)\right).
\end{equation}
Here, $r_{0}$ is the source size parameter.
Therefore, we can obtain the correlation function through assuming 1/4 of the singlet and 3/4 of triplet states. The theoretical correlation function at a given $k^*$ can be calculated as the average FSI weight $\left \langle w\left(\textbf{k}^*,\textbf{r}^*\right) \right \rangle $ obtained from the separation $r^*$, simulated according to the Gaussian law, and the angle between the vectors $\textbf{k}^*$ and $\textbf{r}^*$, simulated according
to a uniform cosine distribution.
The average FSI weight can be described by the formula:
\begin{equation}
w\left(\textbf{k}^*,\textbf{r}^*\right)=\left|
\psi_{-\textbf{k}^*}^{S\left(+\right)}\left(\textbf{r}^*\right)+\left(-1\right)^{S}\psi_{\textbf{k}^*}^{S\left(+\right)}\left(\textbf{r}^*\right)
\right|^{2}/2,
\end{equation}
where $S$ is the total pair spin, $\textbf{r}^*$ is the relative distance, $\psi_{-\textbf{k}^*}^{S\left(+\right)}\left(\textbf{r}^*\right)$ is the equal-time $\left(t^*=0\right)$ reduced Bethe-Salpeter amplitude which can be approximated by the outer solution of the scattering problem \cite{lednicky1982}. This is
\begin{multline}
\psi_{-\textbf{k}^*}^{S\left(+\right)}\left(\textbf{r}^*\right) =e^{i\delta_{c}}\sqrt{A_{c}\left(\lambda \right)} \times\\
\left[e^{-i\textbf{k}^*\textbf{r}^*}F\left(-i\lambda,1,i\xi\right)+f_c\left(k^*\right)\frac{\tilde{G}\left(\rho,\lambda \right)}{r^*}\right],
\end{multline}
where $\delta_{c} =
$arg$\Gamma\left(1+i/k^{*}a_{c}\right)$ is the Coulomb phase corresponding to zero orbital angular momentum, $A_c\left(\lambda \right)=2\pi\lambda \left[\exp\left(2\pi\lambda \right)-1\right]^{-1}$ determines the contribution of the Coulomb interaction, $i.e.$, the positive value corresponding to the repulsion, $\lambda = \left(k^*a_c\right)^{-1}$, $a_c$ = 57.5 $fm$ is the Bohr radius for two protons, $\rho=k^*r^*$,$\xi=\textbf{k}^*\textbf{r}^*+\rho$, $F$ is the confluent hypergeometric function, $\tilde{G}\left(\rho,\lambda \right) = \sqrt{A_{c}\left(\lambda \right)}\left[G_0\left(\rho,\lambda \right)+iF_0\left(\rho,\lambda \right)\right]$ is a combination of the regular $\left(F_0\right)$ and singular $\left(G_0\right)$ s-wave Coulomb function,
\begin{equation}
f_c\left(k^*\right)=\left[\frac{1}{f_0}+\frac{1}{2}d_0k^{*^2}-\frac{2}{a_c}h\left(\lambda \right)-ik^*A_{c}\left(\lambda \right)\right]^{-1}
\end{equation}
is the s-wave scattering amplitude renormalizied by the Coulomb interaction, $d_{0}$ is the effective range of the interaction, and $h\left(\lambda \right) = \lambda^{2}\sum_{n=1}^{\infty}\left[n\left(n^2+\lambda^2\right)\right]^{-1}-C-\ln\left[\lambda \right]$ (here $C$ = 0.5772 is the Euler
constant). The dependence of the scattering parameters on the total pair spin $S$ is
omitted since only the singlet ($S$ = 0) s-wave FSI contributes in the case of identical
nucleons.

\subsection{THE IQMD MODEL}

To apply the above theoretical simulation, the single-particle phase-space distribution at the freeze-out is required. In this work, the correlation function can be established from the emission phase space given by the IQMD transport model \cite{Dan} .

The quantum molecular dynamic model (QMD) is a many-body transport theory, it has been extensively applied to describe heavy-ion reactions from intermediate energy to 2$A$ GeV \cite{Aichelin}. By the QMD studies, various valuable information about both the collision dynamics and the fragmentation process has been learned \cite{Yan,Ma_Shen,Zhang_Ra,Zhou,TaoC,FengZQ,FengZQ2,Huang2,WangTT,Xie}. Excellent extensibility can be also expected due to its microscopic treatment on the collision process. The model mainly consists of several parts: initialization of the projectile and the target nucleons, nucleon transport under  the effective potentials, nucleon-nucleon (NN) binary collisions in a nuclear medium, the Pauli blocking, and the numerical test. The isospin-dependent
quantum dynamic model (IQMD) is based on the QMD model and considers the isospin factors \cite{Pratt} in mean-field, two-body NN collisions, and the Pauli blocking. In the IQMD model, the wave function of each nucleon is represented by the form of Gaussian wave packet, with the parameter $L$ which relates to the size of the reaction system. For Au + Au system, the width $L$ is fixed to 2.16 $fm^{2}$ . Gaussian wave packet is written as
\begin{multline}
\phi_i(\textbf{r}) = \\
\frac{1}{2 \pi L^3/4}\exp\left(\frac{-\left(\textbf{r}-\textbf{r}_{i}\left(t\right)\right)^2}{4L}\right)\exp\left(\frac{i\textbf{r}\cdot\textbf{p}_{i}\left(t\right)}{\hbar}\right).
\end{multline}
Here, $\textbf{r}_{i}\left(t\right)$ and $\textbf{p}_{i}\left(t\right)$
are the time dependent variables which describe the center of the packet in coordinate and momentum space, respectively. Then, all nucleons interact via the effective mean-field and two body NN collisions.

The nuclear mean field can be expressed as:
\begin{equation}
U = U_{Sky}+U_{Coul}+U_{Yuk}+U_{Sym}+U_{MDI}+U_{Pauli},
\end{equation}
where $U_{Sky}$, $U_{Coul}$, $U_{Yuk}$, $U_{Sym}$, $U_{MDI}$,
and $U_{Pauli}$
are the density-dependent Skyrme potential, the Coulomb potential, the surface Yukawa potential, the isospin asymmetry potential, the momentum-dependent interaction and the Pauli potential, respectively. A general review of the above potentials can be found in Ref.~\cite{Aichelin}.
In the present work, the in-medium NN cross section is represented by the formula:
\begin{equation}
\sigma_{NN}^{med} = \left(1-\eta\frac{\rho}{\rho_{0}}\right)\sigma_{NN}^{free},
\end{equation}
where $\rho_{0}$ is the normal nuclear matter density, $\rho$ is the local density, $\eta$ is the in-medium factor and  $\sigma_{NN}^{free}$ is the available experimental NN cross section \cite{Chen1968}. In this above expression, increasing values of parameter $\eta$ correspond to decreasing values of the in-medium nucleon-nucleon cross-section.

\begin{figure*}[h]
 \includegraphics[width=1.\linewidth]{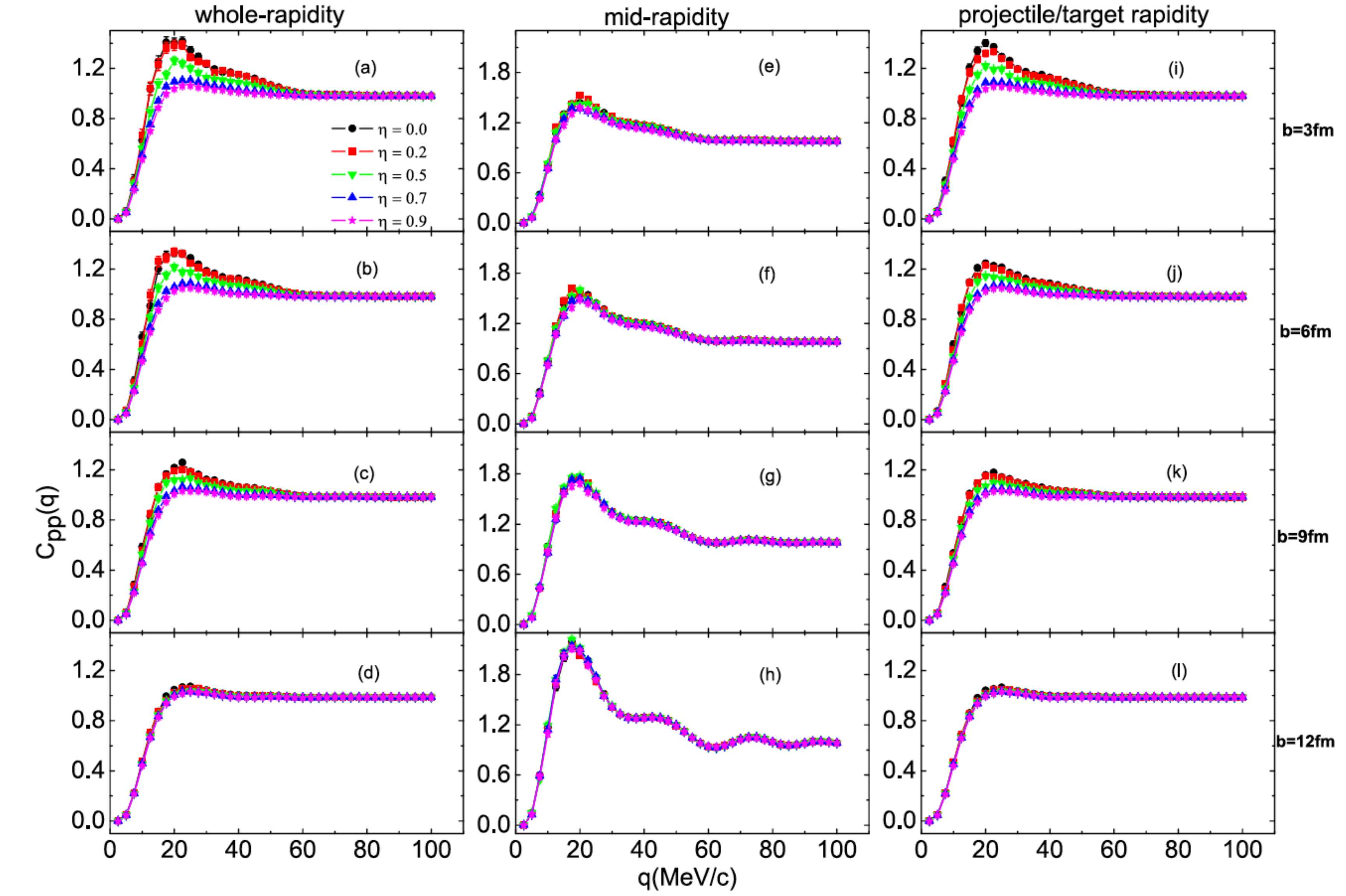}
 \vspace{-0cm}
 \centering
 \caption{(Color online)
Momentum  correlation function of proton-pairs for Au + Au collisions at 1$A$ GeV with different in-medium reduction factors. From top panel to bottom panel, each panel corresponds to impact parameters,
namely b = 3, 6, 9 and 12 fm, respectively. From left to right panel, each panel represents the correlation function constructed for the proton pairs  within whole rapidity, mid-rapidity and projectile/target rapidities, respectively. }
  \label{Fig_Cpp-whole}
 \end{figure*}

In this model, the fragments are identified using a modified minimum spanning tree description. In the Minimum Spanning Tree approach, two nucleons are assumed to share the same cluster if their centers are closer than a distance of 3.5 fm and their relative momentum smaller than 0.3 GeV/c. If nucleon is  not bounded by any clusters, it is  treated by  an emitted (free) nucleon.

\subsection{THE BLAST-WAVE FIT}

In heavy-ion collisions, particles collide with each other randomly, which can be described in term of thermal motion~\cite{Fermi}. We adopt the blast-wave model which has been put forward by Siemens and Rasmussen~\cite{Siemens} to describe the mid-rapidity $p_{T}$ spectra with two free parameters: collective transverse flow velocity $\beta$ and kinetic freeze-out temperature $T_{f}$. The collective transverse flow velocity $\beta$ is parametrized by the surface velocity $\beta_{s}$ in the region of $0 \leq R \leq R_{max}$ \cite{Bondorf}:
\begin{equation}
\beta_r(r) = \beta_{s}\left(\frac{r}{R_{max}}\right)^{\alpha},
\end{equation}
where $R_{max}$ is the maximum radius of the expanding source at thermal freeze-out time, and the $\beta_{s}$ is the particle radial velocity at the maximum surface, e.g. $r= R_{max}$, and the exponent $\alpha$ describes the evolution of the flow velocity with the radius. The $p_{T}$ spectra are a superposition of individual thermal sources with different $r$, which is boosted with the boost angle $\rho=\tanh^{-1}\beta_r\left(r\right)$ ~\cite{Ming2014,Schnedermann}.

\begin{multline}
\frac{dn}{p_{T}dp_{T}}\propto\\
\int_{0}^{R_{max}}rdrm_{T}I_{0}\left(\frac{p_{T}\sinh\rho}{T_{f}}\right)K_{1}\left(\frac{m_{T}\cosh\rho}{T_{f}}\right)
\end{multline}
where, $K_{1}$, $I_{0}$ are the modified Bessel functions. The shapes of the spectra are essentially determined by $T_{f}$, $\beta_{S}$, $\alpha$, and the mass of the particle $m_{0}$. The average flow velocity is estimated by taking an average over the transverse geometry.

\section{ ANALYSIS AND DISCUSSION}

In this work, we use the soft EOS with momentum dependent interaction for  all Au  + Au collisions at  beam energies from 0.4 to 1.5$A$ GeV. The correlation functions are calculated by using the phase-space information from the freeze-out stage.

Firstly, we investigate the influence of the in-medium NN cross section on the momentum correlation function  for Au + Au collisions at 1.0$A$ GeV.
Fig.~\ref{Fig_Cpp-whole} shows the proton-proton momentum  correlation function for Au + Au collisions at 1$A$ GeV with different in-medium reduction factors $\eta$, impact parameters, and proton rapidity region. In each panel, in-medium reduction factor of 0.0, 0.2, 0.5, 0.7 and 0.9 are compared. From top panel to bottom panel, each panel corresponds to different impact parameters, from b = 3, 6, 9 up to 12 fm, respectively. From left to right columns, it represents the correlation function for the proton pairs  within whole rapidity, mid-rapidity and projectile/target rapidities, respectively. Here the mid-rapidity cut means that  both protons are emitted in the rapidity window of -0.5$\leq y/y_{proj.}\leq 0.5$, and  projectile  or target rapidity region means that both protons either come from the rapidity region of  $y/y_{proj.}\geq0.5$ or $y/y_{proj}\leq-0.5 $, where $y$ represents the proton rapidity and $y_{proj.}$ means the initial projectile rapidity.
Overall,  the proton-proton momentum  correlation function  exhibits a peak at relative momentum q = 20 MeV/c, which is due to the strong final-state $s$-wave attraction together with the suppression  at lower relative momentum as a result of Coulomb repulsion and antisymmetrization wave-function  between two protons.

For protons which are emitted in whole rapidity or projectile/target rapidity, the general trend is very similar.
With the increasing of the in-medium NN cross section (i.e. the less in-medium reduction factor $\eta$),
the collision rate between nucleons increases. Therefore more nucleons are emitted early, which make the strength of the momentum correlation function larger. The difference can be further revealed in central and semi-peripheral collisions, however, the difference in momentum correlation function among different $\eta$-factors almost disappears in peripheral collisions. This indicates that the NN cross section in peripheral collisions has no significant change even though the $\eta$ value changes much.
Overall speaking, with the less nucleon-nucleon cross section (i.e. larger $\eta$ factor), the correlation peak decreases, indicating that proton-proton correlation  has a positive correlation with nucleon-nucleon cross section. The sensitivity of correlation strength to the $\eta$ values becomes less important when the reaction goes to peripheral collisions.
 For mid-rapidity protons, correlation functions are much stronger than the cases of whole-rapidity or project/target rapidity, and  shows almost no dependence on in-medium nucleon-nucleon cross section, which indicates very different spacial-time structure of mid-rapidity protons. Essentially, the mid-rapidity protons are emitted very early and non-equilibrium, therefore insensitive to the in-medium NN cross section, additionally they have radial flow which will be mentioned later.
 In addition, another difference is the strength of correlation peak as a function of impact parameter is reverse to the behavior of  correlation peak for whole rapidity or projectile/target rapidity protons, which displays a stronger correlation peak in peripheral collision than central collisions, indicating slighter more compact mid-rapidity source in peripheral collisions.

\begin{figure*}[htbp]
 \includegraphics[width=1.\linewidth]{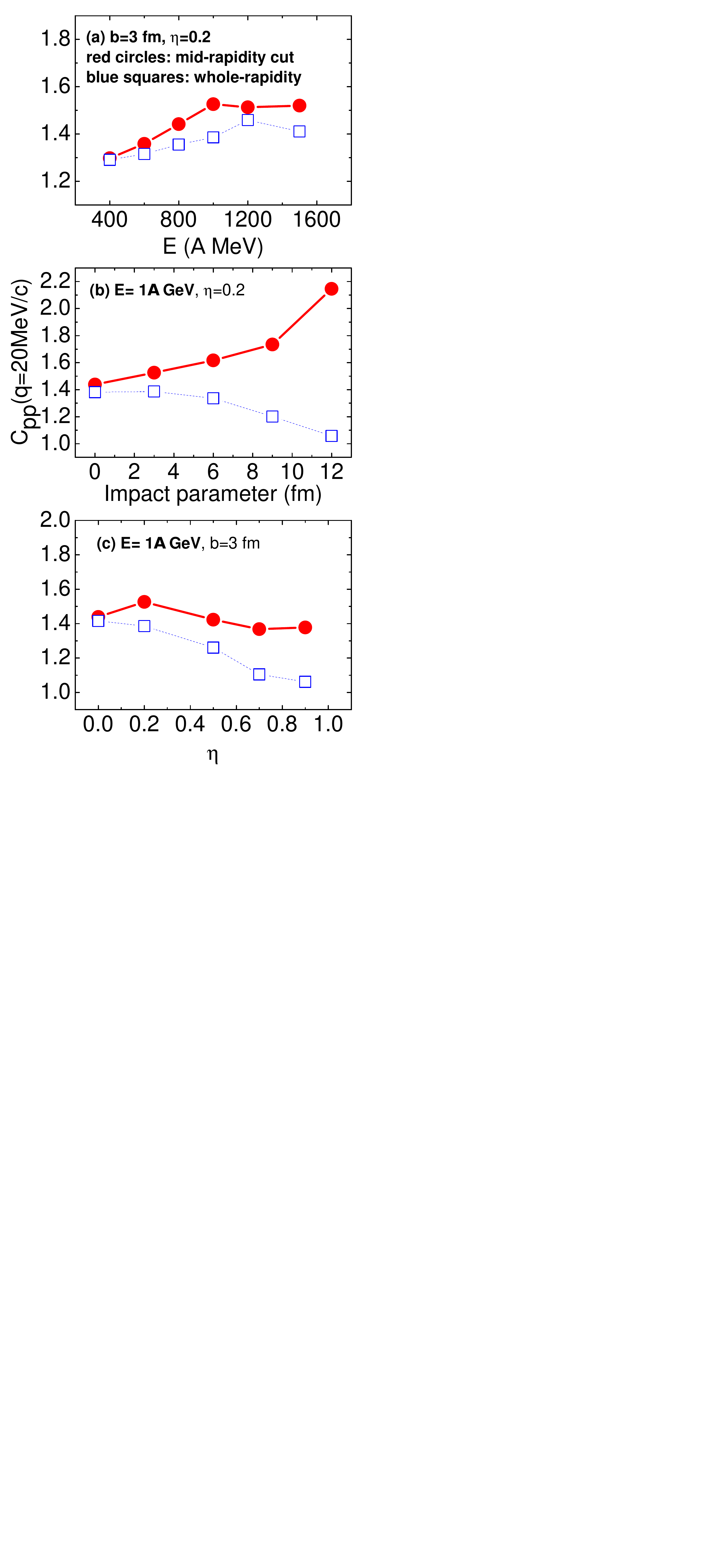}
 \vspace{-20cm}
 \centering
 \caption{(Color online)
 $\left(a\right)$ $C_{pp}$  strength at 20 MeV/c as a function of the incident energy, which is calculated at an impact parameter of b = 3 fm and in-medium reduction factor of $\eta$ = 0.2;
   $\left(b\right)$ $C_{pp}$ strength at 20 MeV/c as a function of the impact parameter, which is calculated at 1$A$ GeV and the in-medium reduction factor $\eta$ = 0.2;
   $\left(c\right)$ $C_{pp}$ strength at 20 MeV/c as a function of the in-medium reduction factor, which is calculated at  b = 3 fm and 1$A$ GeV. Note that the red circles corresponds to mid-rapidity cut and the blue squares whole-rapidity cut.
 }
 \label{Cpp_Eb-eta}
\end{figure*}

\begin{figure*}[h]
\includegraphics[width=1.\linewidth]{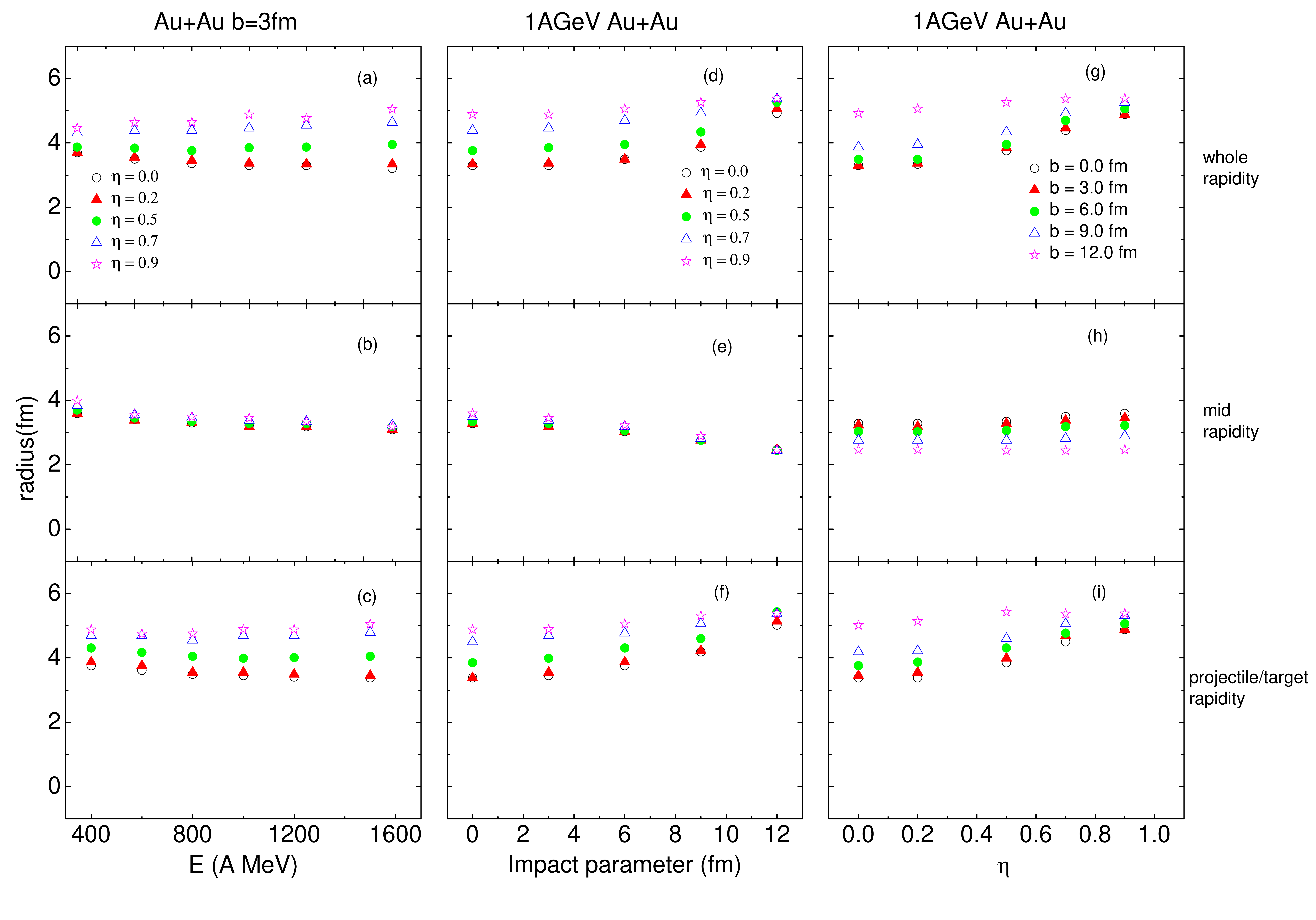}
\caption{(Color online)
Gaussian source radius as a function of the incident energy in different $\eta$-factor at fixed impact parameter b = 3 fm (left column), as a function of impact parameter
with different $\eta$ factors (middle column), as a function of in-medium nucleon-nucleon cross section reduction factor at different impact parameter (right column). From upper to lower rows, it corresponds to  the Gaussian radius of free protons emitted from  the whole rapidity window,  the mid-rapidity window, and the target or projectile rapidity window, respectively.}
\label{Fig_radius-E}
\end{figure*}

\begin{figure*}[h]
\includegraphics[width=\linewidth]{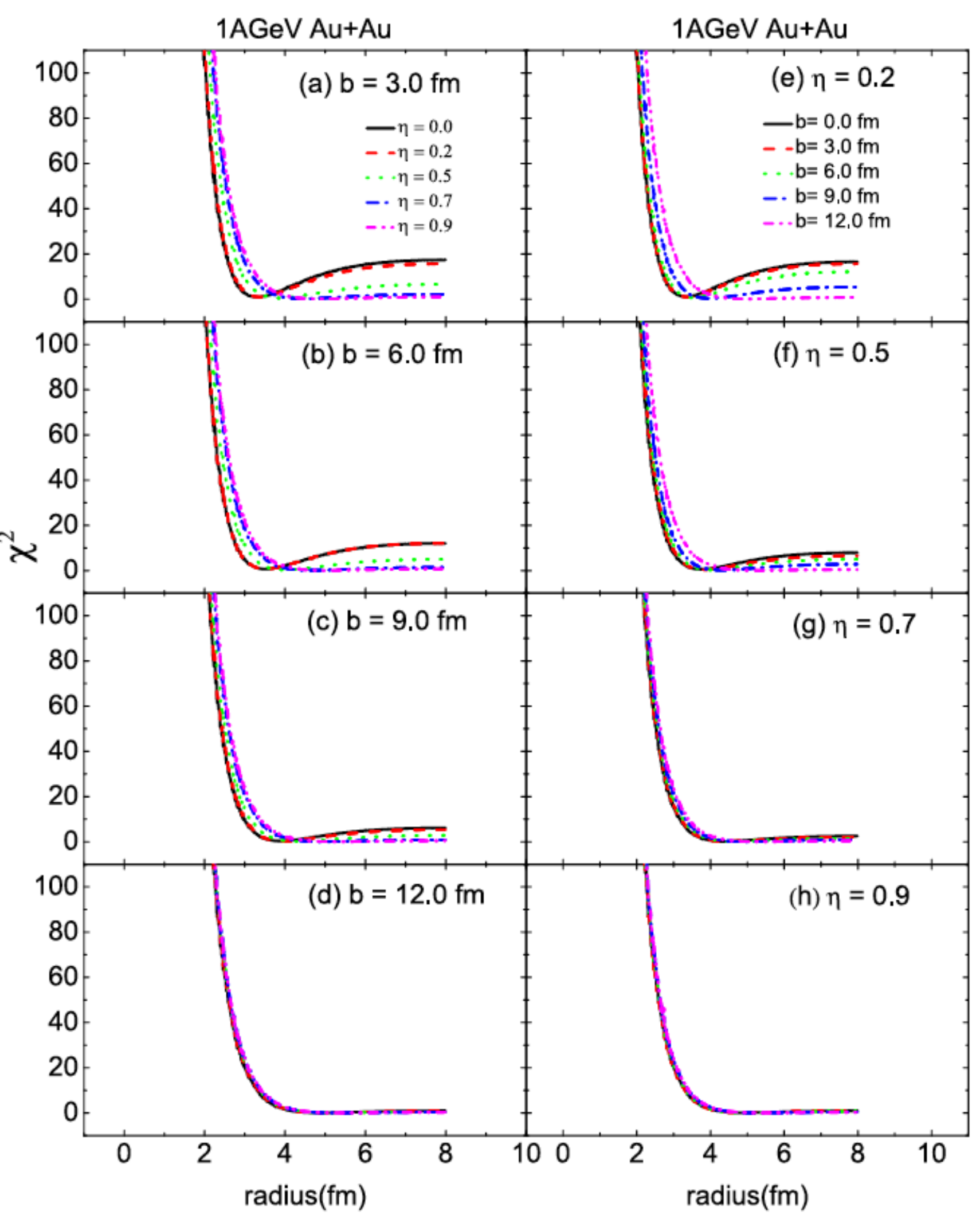}
\caption{(Color online)
The $\chi$-square obtained from the fits of the proton-proton momentum correlation function by the Lednicky et al. analytical formalism calculation as a function of the radius of the Gaussian source. Left panels: each panel represents different fixed impact parameters with different $\eta$ factor; Right panel: each panels represent different fixed  $\eta$ factors at different impact parameters, respectively.}
\label{Fig_chis2}
\end{figure*}

\begin{figure*}[h]
 \includegraphics[width=1\linewidth]{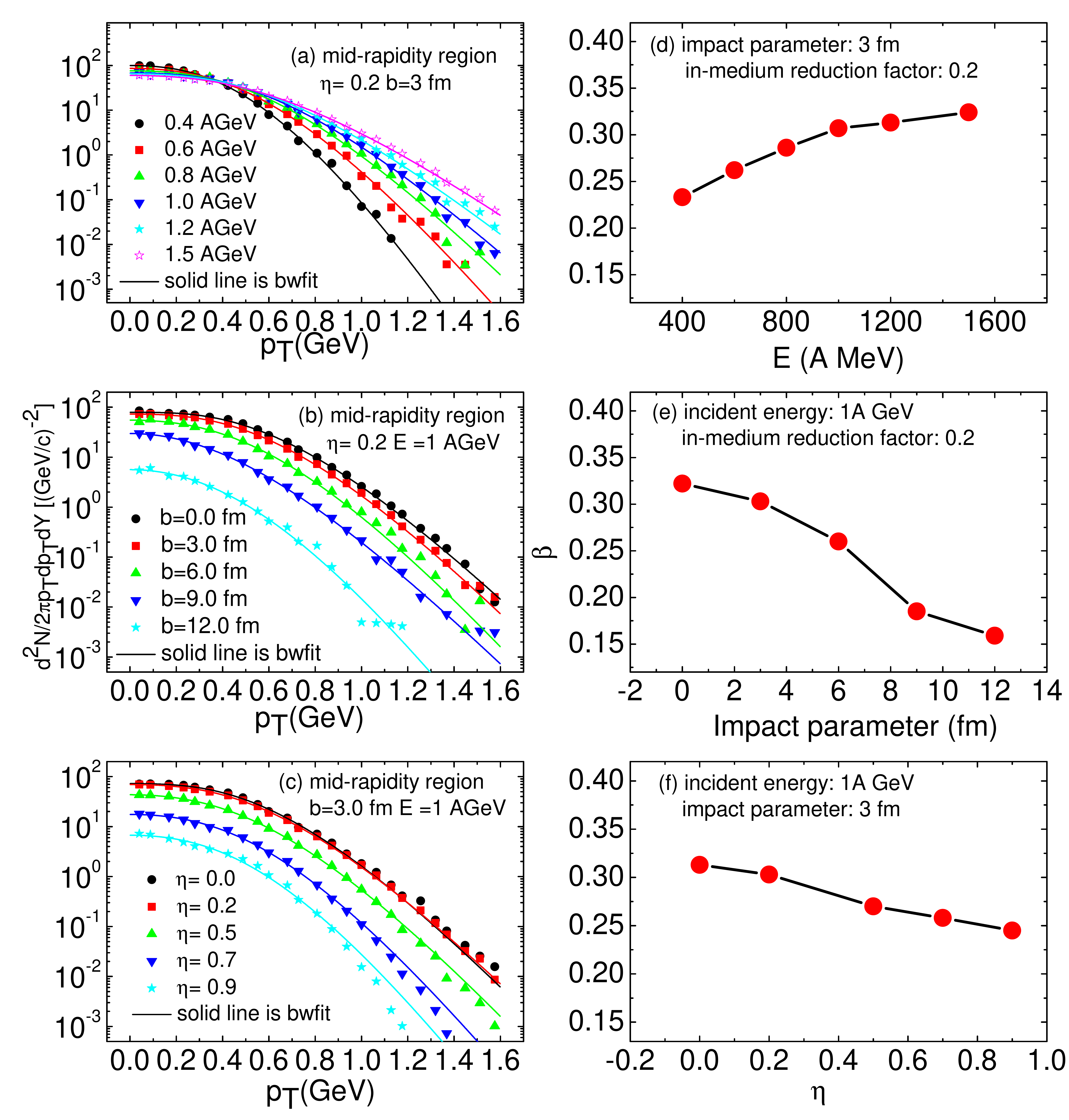}
 \centering
 \caption{(Color online)
  Left panels: Blast-wave fitting to $p_{T}$ spectra of protons in the mid-rapidity region. upper panel:  incident energy dependence for b=3 fm and $\eta$=0.2; middle panel: impact parameter dependence at E = 1$A$ GeV and $\eta$ = 0.2; lower panel: in-medium cross section dependence at b = 3 fm and 1$A$ GeV.
  Right panels: the fitted radial flow parameter ($\beta$) as a function of beam energy (upper panel), impact parameter at 1$A$ GeV and $\eta$ = 0.2 (middle panel), in-medium cross section reduction factor (lower panel) at 1$A$ GeV and b = 3 fm.
 }
 \label{Fig_BWfits}
\end{figure*}

How the three variables including the incident energy, impact parameter and  in-medium nucleon-nucleon cross section affect the strength of $C_{pp}$ is presented in Fig.~\ref{Cpp_Eb-eta} for proton-proton correlation function where the whole rapidity window (blue squares) or mid-rapidity (solid circles) is applied for emitted protons  in  Au + Au collisions. For the incident energy dependence,
the increase of  momentum correlation peak from low energy to high energy is cleanly seen in panel $\left(a\right)$ of Fig.~\ref{Cpp_Eb-eta} , and the mid-rapidity cut displays a stronger peak. It can be generally understood that  a more rapid collision process, and a smaller emission source space and short time interval among emitting nucleons at higher energy, so does the mid-rapidity source \cite{Ma2006}. In the panel $\left(b\right)$ of Fig.~\ref{Cpp_Eb-eta} , we see that the correlation strength increases with the impact parameter for the mid-rapidity cut but opposite for the whole rapidity cut, which shows that geometrical cut for two rapidities are complemented, e.g. stronger correlation  for mid-rapidity protons at peripheral collision indicates smaller source size, but weaker  correlation  for whole-rapidity protons indicates larger source size.
In the panel $\left(c\right)$ of Fig.~\ref{Cpp_Eb-eta} , with the increase of the in-medium cross section modification factor, $i.e.$ decreasing the in-medium NN cross section, the peak strength becomes smaller. In the other words, proton-proton correlation positively depends on the nucleon-nucleon collisions.
On the other hand, in contrast with the  whole-rapidity protons, mid-rapidity protons show weaker sensitivity to the NNCS.
As we  have already known that the strength of the correlation function depends mainly on the source size,  the above  behavior of the HBT strength essentially reflect the changing size of the emission source versus the beam energy, impact parameter and nucleon-nucleon cross section.

Fig.~\ref{Fig_radius-E} presents the radius of the Gaussian source for emitted protons as a function of beam energy (left column), impact parameter (middle column) and in-medium cross section reduction factor (right column) for different rapidity windows, namely the whole rapidity (upper row), mid-rapidity (middle row) and projectile/target rapidity (bottom row). Overall,  for mid-rapidity proton-proton correlations, the sensitivity to the in-medium cross section reduction factor is almost not visible.  The source radius shows  a slight drop with the the increasing of beam energy or impact parameter.
For the whole rapidity or projectile/target rapidity windows, their beam energy dependences are very similar if the same $\eta$ is applied, and source radius  increases with the increasing of $\eta$, i.e. the decreasing of nucleon-nucleon cross section. When the nucleon-nucleon cross section is larger (e.g. $\eta \leq 0.5$), source radius drops with the incident energy, indicating fast emission and/or compact proton emission size in higher energies. However, the situation is different when the nucleon-nucleon cross section is small  (e.g. $\eta > 0.7$), where radius shows a slight increasing or a plateau behaviour.  But overall, even though slight  beam energy dependence of radius is seen, the trend is rather weak as seen  in ultra-relativistic energy heavy ion collisions \cite{ZhangS}.

Middle  column and right column demonstrate  the radius of the Gaussian source as a function of impact parameter at different fixed $\eta$ values  or $\eta$ at  different impact parameters, respectively.
As expected, for correlation between protons from the whole rapidity region or projectile/target region, the source size increases with the increasing of impact parameter, and the larger the nucleon-nucleon cross section, the stronger the dependence of the source size on impact parameter. With increasing impact parameters, the effective source size gets bigger because apparently the (target or projectile) spectator region gets bigger and bigger with increasing impact parameter, and this effect is more pronounced if the in-medium decrease of the cross-section is smaller. This indicates some geometrical effect and protons coming from spectator fragmentation mechanism. At the same time,  the source size increases with the decreasing of the in-medium NN cross section (i.e. larger $\eta$ factor). The smaller the impact parameter, the stronger the dependence of the source size on in-medium NN cross section, which can be understood that more frequent collision effect is expected to affect the dynamical evolution.
However,  for correlation between protons from the mid-rapidity region, the tendency of the source size becomes decreasing with the increasing of impact parameter, regardless of the in-medium nucleon-nucleon cross section.  This reverse dependence of source radius  between mid-rapidity protons and projectile/target rapidity protons as a function of impact parameter  indicates some geometrical evolution of participant  and spectator region, which can be also well seen from Fig.~\ref{Cpp_Eb-eta}(b).

We should mention that to extract the above source size, theoretical calculations for $C_{pp}$ was performed by using the $Lednick\acute{y}$ and $Lyuboshitz$ analytical method.  The best fitting source size is judged by finding the minimum of the reduced chi-square. Fig.~\ref{Fig_chis2}  present examples of the $\chi^2$-variance between the IQMD calculations with the $Lednick\acute{y}$-$Lyuboshitz$ analytical formalism and the Gaussian source correlation as a function of the radius of the Gaussian source in different impact parameters with the changing $\eta$ factors (left panels) or in different $\eta$ factors with the changing impact parameters (right panels). Generally, the minimum can be well defined, but the errors on its location are apparently rather asymmetric.

Middle rapidity protons are possibly experienced by the collective radial flow expansion in comparison with the projectile/target rapidity protons.
To demonstrate how large the radial flow is for the mid-rapidity protons,  we use the Blast-wave (BW) fits  to $p_{T}$ spectra of protons in the mid-rapidity region. Here, we get the $\alpha$ value equal to $1/3$ on the $p_{T}$ spectra of central collisions and fix this value when do the fitting on the other impact parameters \cite{Ming2014}.
Left column  of Fig~\ref{Fig_BWfits} shows the $p_{T}$ spectra of mid-rapidity protons in different incident energy for b = 3 fm and $\eta$ = 0.2 (top row), in different impact parameter for  E = 1$A$ GeV and $\eta$ = 0.2 (middle row) and with different in-medium cross section for b = 3 fm and 1$A$ GeV (bottom row), where the solid symbols represent the calculated results from the IQMD and the solid lines are the BW fits. Overall all lines can well reproduce the spectra, from which the radial flow parameters can be systematically extracted.
Right column of Fig.~\ref{Fig_BWfits}  displays the extracted radial flow parameters ($\beta$) as a function of incident energy at b = 3 fm and $\eta$ = 0.2 (top panel), as a function of impact parameter at E = 1$A$ GeV and  $\eta$ = 0.2 (middle panel), and as a function  of in-medium cross section factor at   E = 1$A$ GeV and b = 3 fm (bottom panel). Obviously, the radial flow becomes stronger in higher incident energy as well as  in more central collisions. Meanwhile,  larger in-medium nucleon-nucleon cross-section (i.e. smaller $\eta$ values) leads to larger radial flow due to the frequent nucleon-nucleon collision in overlap zone. It is noticed that radial flow has been already extensively discussed in mediate and high energy HIC, eg. discussed in  Ref.~\cite{Helgesson} by Helgesson {\it et al.}.

\begin{figure*}[h]
 \includegraphics[width=0.5\linewidth]{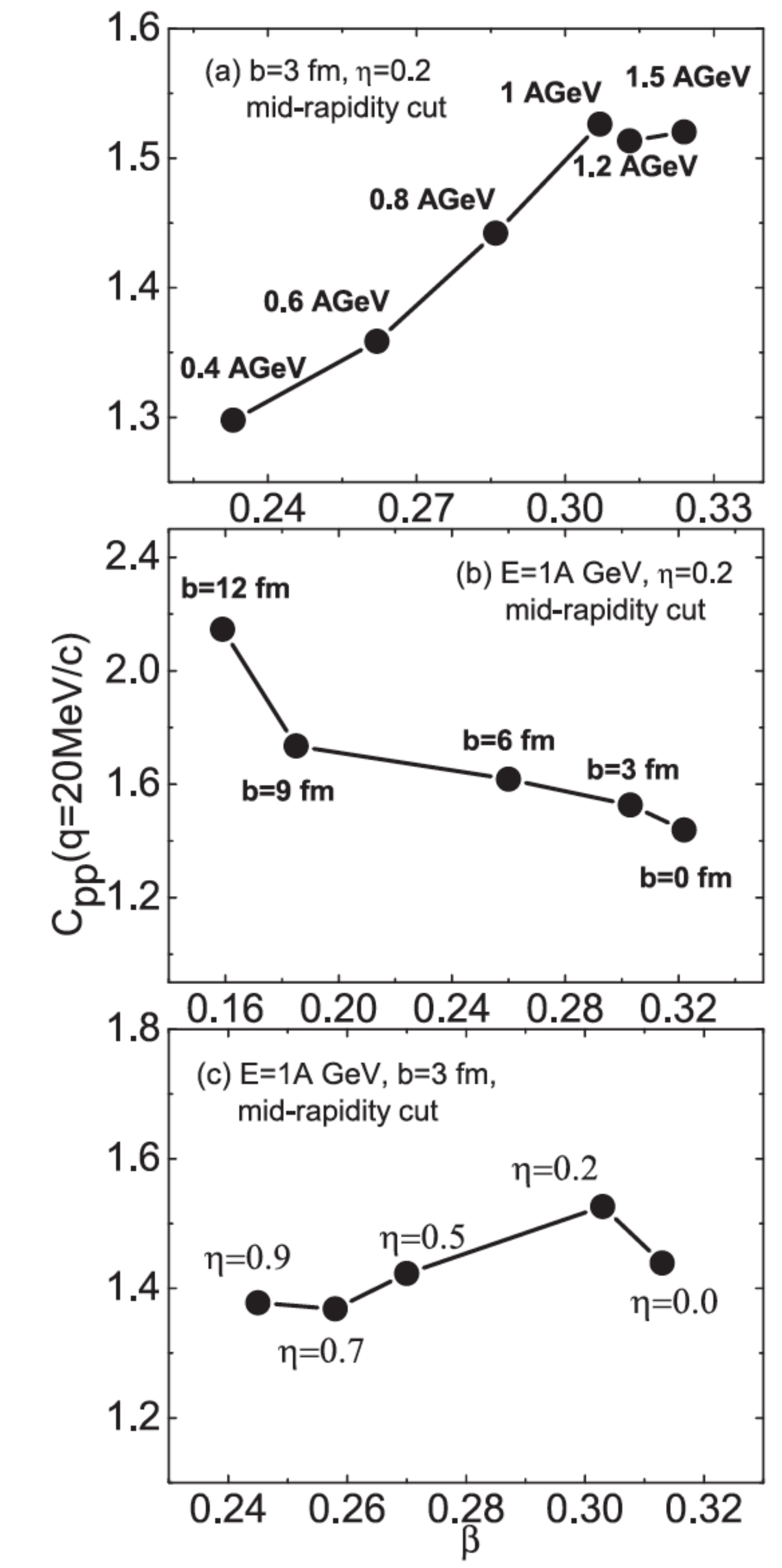}
 \centering
 \caption{ Correlation strength at $q$ = 20 MeV/c as a function of the radial flow parameter ($\beta$) in different conditions. Upper row: beam energy is a variable as shown in figure, at fixed b = 3 fm and $\eta$ = 0.2 (upper row); Middle row: impact parameter is a variable as shown in figure, at fixed E = 1$A$ GeV and $\eta$ = 0.2; Bottom row: the in-medium nucleon-nucleon cross section reduction factor is a variable as shown in figure, at fixed E = 1$A$ GeV and b = 3 fm.
 }
 \label{Fig_Cpp-beta}
\end{figure*}

 For mid-rapidity p-p correlation, let us have a close look for correlation strength versus the radial flow. Since we have relationship between the correlation strength at 20 MeV/c ($C_{pp}$(q=20MeV/c) ) versus beam energy as well as the radial flow ($\beta$) versus beam energy, we can obtain the relationship between $C_{pp}$(q=20MeV/c) and $\beta$ which is displayed in Fig.~\ref{Fig_Cpp-beta}(a) in the condition of b = 3 fm and $\eta$ = 0.2. It tells us that the larger the radial flow velocity, the stronger the proton-proton correlation.
In the same way, we got  $C_{pp}$(q=20MeV/c) versus $\beta$ in Fig.~\ref{Fig_Cpp-beta}(b) in the condition of  E = 1$A$ GeV and $\eta$ = 0.2 where the impact parameter is a variable. In this case, anti-correlation is observed. Simiarily,  $C_{pp}$(q=20MeV/c) versus $\beta$ is shown
in Fig.~\ref{Fig_Cpp-beta}(c) in the condition of E=1$A$ GeV and b = 3 fm where the in-medium nucleon-nucleon cross section reduction factor is a variable. Here, a slight increasing behavior  is demonstrated.  Seen from the above three variables, we found no unique dependence of  p-p correlation function as a function of radial flow parameter.

\section{SUMMARY}

In the present work, we use the IQMD transport approach to calculate the phase-space points at the freeze-out stage for Au + Au collisions from 0.4 to 1.5$A$ GeV. Afterwards the phase space points
were processed within the $Lednick\acute{y}$ and $Lyuboshitz$ analytical formalism to reconstruct the proton-proton correlation function.
In this way, we systematically study how the in-medium $NN$ cross section  affects the strength of the momentum correlation function of proton-proton pairs for Au + Au from 0.4 to 1.5$A$ GeV in different rapidity windows and different impact parameters.
Results show that the larger in-medium $NN$
cross section results in a stronger momentum correlation function than a smaller  in-medium $NN$
cross section, especially at small impact parameters, for the whole rapidity or projectile/target  rapidity proton pairs. This behavior is interpreted as stronger correlation for equilibrium-like protons induced by the higher nucleon-nucleon collision rate. However, for the mid-rapidity proton emission, the in-medium nucleon-nucleon cross section has less effect on the momentum correlation function due to very different emission mechanism.
In addition, the impact parameter effect on the HBT strength has been also addressed in the work. We have shown that the HBT strength has very different dependence between the whole rapidity or projectile/target  rapidity proton pairs and the mid-rapidity  proton pairs.
For  proton pairs from  projectile/target  rapidity, it decreases
with the increasing of impact parameter; but for proton pairs from  mid-rapidity,  it increases
with the increasing of impact parameter.
By fitting the momentum correlation function with the Gaussian source, the effective proton  emission source sizes are extracted.
Results show that the source radius general increases with the increasing of the impact parameter for emitted protons within the whole rapidity or projectile (target) rapidity, however,  it decreases versus the impact parameter for the mid-rapidity protons and it shows insensitivity to the in-medium nucleon-nucleon cross section. The above phenomenon  reflects the evolution of source size with the collision geometry, i.e. the mid-rapidity source becomes smaller but the projectile/target source becomes larger with the increasing of impact parameter.
Moreover, the beam energy dependence of HBT strength is also presented. Generally, the HBT strength shows a slight change with beam energy especially for  the whole rapidity or projectile (target) rapidity proton pairs.
By using the Blast-wave fits to the transverse momentum spectra  of mid-rapidity protons, the radial flow parameters are systemically extracted as a function of beam energy, impact parameter and in-medium nucleon-nucleon reduction factor, and therefore relationships of HBT strength versus radial flow parameter are constructed in different conditions. However, no unique dependence is found, which indicates the radial flow is not a decisive variable for the p-p correlation.

\begin{acknowledgments}
 This work was supported partly by the National Natural Science Foundation of China under Contract Nos. 11421505, 11220101005, the Major State Basic Research Development Program in China under Contract  2014CB845401, and the Key Research Program of Frontier Sciences of the CAS under Grant No. QYZDJ-SSW-SLH002.
\end{acknowledgments}

\end{document}